\documentclass[twocolumn]{aastex62}

\usepackage{CJK}

\newcommand{\kms}{km\,s$^{-1}$} 
 
\newcommand{\lsun}{$L_\odot$}
\newcommand{\msun}{$M_\odot$}
\newcommand{\ujb}{$\mu$Jy\,beam$^{-1}$}
\newcommand{\mjb}{mJy\,beam$^{-1}$}
\newcommand{\jyb}{Jy\,beam$^{-1}$}
\newcommand{\myalpha}{$\alpha_{\rm 0.9\,cm-1.3\,mm}$}
\newcommand{\MMW}{MM2W} 
\newcommand{\MME}{MM2E}

\shorttitle{Protobinary in  G11.92$-$0.61 MM2}
\shortauthors{Cyganowski et al.}

\begin{document}
\begin{CJK*}{UTF8}{gbsn}
\title{Discovery of a 500 au Protobinary in the Massive Prestellar Core G11.92$-$0.61 MM2}

\correspondingauthor{C. J. Cyganowski}
\email{cc243@st-andrews.ac.uk}

\author[0000-0001-6725-1734]{C. J. Cyganowski}
\affil{SUPA, School of Physics and Astronomy, University of St. Andrews, North Haugh, St. Andrews KY16 9SS, UK}

\author[0000-0003-1008-1142]{J. D. Ilee}
\affil{School of Physics and Astronomy, University of Leeds, Leeds LS2 9JT, UK}

\author[0000-0002-6558-7653]{C. L. Brogan}
\affil{National Radio Astronomy Observatory, 520 Edgemont Rd, Charlottesville, VA 22903, USA}

\author[0000-0001-6492-0090]{T. R. Hunter}
\affil{National Radio Astronomy Observatory, 520 Edgemont Rd, Charlottesville, VA 22903, USA}
\affil{Center for Astrophysics $\mid$ Harvard \& Smithsonian, Cambridge, MA 02138, USA}

\author[0000-0002-8389-6695]{Suinan Zhang (张遂楠)}
\affil{SUPA, School of Physics and Astronomy, University of St. Andrews, North Haugh, St. Andrews KY16 9SS, UK}

\author[0000-0001-8228-9503]{T. J. Harries}
\affil{Department of Physics and Astronomy, University of Exeter, Stocker Road,  Exeter EX4 4QL, UK}

\author[0000-0002-9593-7618]{T. J. Haworth}
\affil{Astronomy Unit, School of Physics and Astronomy, Queen Mary University of London, London E1 4NS, UK}



\begin{abstract}

We present high-resolution ($\lesssim$160\,au) Atacama Large Millimeter/submillimeter Array (ALMA) 1.3~mm observations of the high-mass prestellar core candidate G11.92$-$0.61 MM2, which reveal that this source is in fact a protobinary system with a projected separation of 505\,au.  The binary components, \MME\/ and \MMW\/, are compact (radii$<$140\,au) sources within the partially optically thick dust emission with $\alpha_{\rm 0.9\,cm-1.3\,mm}=$2.47--2.94.
The 1.3~mm brightness temperatures, T$_{\rm b}=$68.4/64.6\,K for \MME/\MMW, imply internal heating and minimum luminosities L$_*>$24.7\,\lsun\/ for \MME\/ and L$_*>$12.6\,\lsun\/ for \MMW.  The compact sources are connected by a ``bridge'' of lower-surface-brightness dust emission and 
lie within more extended emission 
that may correspond to a circumbinary disk.
The circumprotostellar gas mass, estimated from $\sim$0\farcs2-resolution VLA 0.9\,cm observations assuming optically thin emission, is 6.8$\pm$0.9 \msun.
No line emission is detected towards \MME\/ and \MMW\/
in our high-resolution 1.3\,mm ALMA observations.  The only line detected is $^{13}$CO J=2-1, in absorption against the 1.3\,mm continuum, 
which likely traces a layer of cooler molecular material surrounding the protostars.
We also report the discovery of a highly asymmetric bipolar molecular outflow that appears to be driven by \MME\/ and/or \MMW\/ in new deep, $\sim$0\farcs5-resolution (1680\,au) ALMA 0.82\,mm observations.
This outflow, traced by low-excitation CH$_3$OH emission, indicates ongoing accretion onto the protobinary system.
Overall, the super-Alfv\'{e}nic models of \citet{mignon-risse21} agree well with  the observed properties of the \MME/\MMW\/ protobinary, suggesting that this system may be forming in an environment with a weak magnetic field.

\end{abstract}


\section{Introduction} \label{sec:intro}

Binarity and multiplicity are conspicuous characteristics of main-sequence O and early-B type stars \citep[e.g.][]{chini12,sana14,gravity18} that must be explained by models of high-mass star formation.
While recent observational advances have revealed binaries in
massive young stellar objects \citep[MYSOs; e.g.][]{beltran16,beuther17,kraus17,pomohaci19,zapata19,zhang19,tanaka20,koumpia21}, all of these sources are already infrared-bright and/or evolved enough for the binary components to excite hypercompact (HC) or ultracompact (UC) HII regions.
There thus remains a lack of observational evidence for the earliest stages of high-mass binary formation.

These early stages, however, are important for constraining models of high-mass star formation, which differ in their predictions for the formation pathways and mass ratios of young binary or multiple systems.  Modelling the collapse of isolated massive prestellar cores including turbulence and radiative and outflow feedback, \citet{rosen20} find that companion stars form via turbulent fragmentation at early times and via disk fragmentation at late times (while with a strong magnetic field, no companion stars are formed). 
Other recent magnetohydronamic \citep[e.g.][]{mignon-risse21} and hydrodynamic \citep[e.g.][]{oliva20} models of the collapse of massive cores indicate 
that binaries form via disk rather than core fragmentation, with disk spiral arms playing an important role.  
While the binary formed in the early radiation-hydrodynamic simulations of \citet{krumholz09} consists of two high-mass stars, many subsequent works \citep[e.g.][]{rosen16,rosen19,rosen20,meyer18} instead predict the formation of hierarchical systems with a single high-mass member. 
Notable recent exceptions are the super-Alfv\'{e}nic cases of \citet{mignon-risse21}, which form stable binary systems with mass ratios of $\approx$1.1-1.6 and separations of a few hundred au.

Observationally, massive prestellar cores such as those adopted as initial conditions in the aforementioned simulations have proven elusive \citep[e.g.][and references therein]{redaelli21}.  Among the longest-standing candidates is G11.92$-$0.61 MM2 (hereafter MM2): the second-brightest millimeter continuum core in the G11.92$-$0.61 protocluster \citep{cyganowski11a,cyganowski17}. MM2 was identified as a candidate massive prestellar core based on its lack of molecular line emission and other star formation indicators in Submillimeter Array (SMA) and Karl G. Jansky Very Large Array (VLA) observations \citep{cyganowski14}.  
MM2 is only $\sim$7\farcs2 (0.12\,pc) from G11.92$-$0.61-MM1, a proto-O star with a fragmented Keplerian disk \citep{ilee16,ilee18}; as in \citet{cyganowski14,cyganowski17}, here we adopt $d_{\rm MM2}$=3.37$^{+0.39}_{-0.32}$ kpc, the maser parallax distance for MM1 \citep{sato14}.
From the SMA dust continuum, \citet{cyganowski14} estimated that MM2's mass is M$\gtrsim$30 M$_{\odot}$ within a radius $<$1000\,au.

In this Letter, we present the serendipitous discovery that MM2 is a candidate (proto)binary in new high-resolution ($\lesssim$0\farcs05, $\lesssim$160\,au) 1.3\,mm Atacama Large Millimeter/Submillimeter Array (ALMA) observations targeting the MM1 disk.  To better understand the properties and evolutionary states of the binary components, we complement these data with VLA 3 and 0.9\,cm continuum observations (resolution $\sim$0\farcs2$\sim$700\,au) and lower-resolution ($\sim$0\farcs5$\sim$1700\,au) ALMA 0.82 and 1.05\,mm observations.

\begin{deluxetable*}{lcccccc}
\tablewidth{0pc}
\setlength{\tabcolsep}{0.5mm}
\tablecaption{Observational and image parameters\label{tab:obstable}}  
\tablehead{ & \multicolumn{4}{c}{ALMA} & \multicolumn{2}{c}{VLA}\\
\cline{2-5} 
\colhead{Parameter} & \colhead{Cycle 6} & \colhead{Cycle 4} & \colhead{Cycle 2} & \colhead{Cycles 3-5} & &
}
\startdata
Wavelength & \colhead{1.3~mm} & \colhead{1.3~mm} & \colhead{1.05~mm} & \colhead{0.82~mm}  & {3~cm} & {0.9~cm} \\
Observing date(s) (UT) & 2019 Jul 15-16  & 2017 Aug 7-9 & 2015 May 14 & 2018 Jul 10, Aug 16 & 2015 Jun 25 & 2015 Feb 9-10 \\
 & & & & 2017 Apr 22, 26 & & \\
  & & & & 2016 Apr 9 & & \\
Project Code(s) & 2018.1.01010.S & 2016.1.01147.S & 2013.1.00812.S & 2015.1.00827.S,  & 15A-232 & 15A-232 \\
 & & & & 2017.1.01373.S  & & \\
Configuration(s) & C43-8 & C40-7 & C34-3(4)& C43-1, C43-2, & A & B \\
 & & & & C40-3, C36-2/3 & & \\
Number(s) of antennas & 42 & 45 & 37 & 41-46 & 27 & 26-27 \\
Phase Center (J2000): & & & & & &  \\
R.A. &  18$^{\rm h}$13$^{\rm m}$58$^{\rm s}$.1099 & 18$^{\rm h}$13$^{\rm m}$58$^{\rm s}$.1099 & 18$^{\rm h}$13$^{\rm m}$58$^{\rm s}$.110\tablenotemark{a} & 18$^{\rm h}$13$^{\rm m}$57$^{\rm s}$.8599 &  18$^{\rm h}$13$^{\rm m}$58$^{\rm s}$.10 & 18$^{\rm h}$13$^{\rm m}$58$^{\rm s}$.10 \\
Dec. & $-$18$^{\circ}$54\arcmin20\farcs141&  $-$18$^{\circ}$54\arcmin20\farcs141 & $-$18$^{\circ}$54\arcmin22\farcs141\tablenotemark{a} & $-$18$^{\circ}$54\arcmin13\farcs958 &$-$18$^{\circ}$54\arcmin16\farcs7 & $-$18$^{\circ}$54\arcmin16\farcs7 \\
Primary beam (FWHP) & 26\arcsec & 26\arcsec & mosaic  & 17\arcsec & 4\arcmin & 1.3\arcmin \\
Frequency coverage\tablenotemark{b}: & & & & & & \\
Lower band (LSB)  & 220.530~GHz   & 220.530~GHz & 278.23 GHz & 358.02 GHz & 9 GHz & 31 GHz \\
center(s) &  221.500~GHz & 221.500~GHz & &  & & \\
Upper band (USB)  &  235.780~GHz  & 235.780~GHz & 290.62 GHz & & 11 GHz & 35 GHz\\
center(s) &  238.850~GHz & 238.850~GHz & 292.03 GHz &  & & \\
Bandwidth(s)\tablenotemark{b} & 4$\times$937.5 MHz & 1$\times$468.75 MHz & 2$\times$1.875 GHz & 1.875 GHz & 2$\times$2.048 GHz & 4$\times$2.048 GHz\\
 & & 3$\times$937.5 MHz & 117.2 MHz & & & \\
Channel spacing(s)\tablenotemark{b} & 0.244~MHz & 0.122~MHz & 0.977 MHz & 0.977 MHz & 1 kHz & 1 kHz \\
 & &  0.244~MHz  & 0.122 MHz & & & \\
  & &  0.488~MHz  & & & & \\
Gain calibrator(s) & J1832-2039  & J1832-2039 & J1733-1304 & J1911-2006, J1733-1304 & J1832-2039 & J1832-2039\\
Bandpass calibrator & J1924-2914 & J1924-2914 &  J1733-1304 & J1924-2914 & J1924-2914 & J1924-2914\\
Flux calibrator(s) & J1924-2914 &  J1733-1304   & Titan  & J1924-2914, Titan & J1331$+$3030 & J1331$+$3030 \\
Projected baselines (k$\lambda$) & 84--6298  & 14--2787 & 20--528 & 14--583 & 17--1221 & 9--1225\\
Largest angular scale (LAS)\tablenotemark{c} & 0\farcs8 & 1\farcs4 & 4\farcs2 & 4\farcs5 & 3\farcs4 & 4\farcs2 \\
Reference(s)\tablenotemark{d} & \nodata & I18 & C17 & \nodata & I16,C17 & I16,C17 \\
Robust parameter (R) & \multicolumn{2}{c}{various (as indicated below)} & 0.5 & 0.5 & 0.5 & 0.0 \\
Synthesized beam\tablenotemark{e} & \multicolumn{2}{c}{R=+1: 104$\times$81 [$-$82$^{\circ}$]}  & 534$\times$387 [$-$83$^{\circ}$] & 574$\times$433 [$-$80$^{\circ}$] & 298$\times$168 [0$^{\circ}$] &    270$\times$144 [$-$6$^{\circ}$]\\
(mas$\times$mas[PA]) & \multicolumn{2}{c}{R=0: 57$\times$41[+64$^{\circ}$]} & & & \\
& \multicolumn{2}{c}{R=$-1$: 40$\times$32 [+68$^{\circ}$]} & & & \\
\multicolumn{2}{l}{Rms noise\tablenotemark{f} (\mjb):}   & & & &  & \\
Continuum  & \multicolumn{2}{c}{R=+1: 0.034 } &   \nodata & \nodata & 0.0056 & 0.0084\\
 & \multicolumn{2}{c}{R=0: 0.038 } & & & \\
& \multicolumn{2}{c}{R=$-1$: 0.074 } & & & \\
Spectral line  & \multicolumn{2}{c}{R=0.5: 0.68} & 4.9 ($\Delta$v=1.0 km s$^{-1}$) & 1.2 & \nodata & \nodata \\
& \multicolumn{2}{c}{R=0 ($^{13}$CO): 0.91} & 2.8-3.5 ($\Delta$v=1.2 km s$^{-1}$) & &  & \\
\enddata
\tablenotetext{a}{Central pointing}
\tablenotetext{b}{ALMA 1.05 and 0.82~mm: details only for spw(s) containing lines discussed in this Letter.  The narrow 1.05~mm spw targeted H$_2$CO 4$_{0,4}-$3$_{0,3}$ at 290.62341 GHz.  Band centers: ALMA: rest frequency,  VLA: sky frequency. } 
\tablenotetext{c}{Estimated using the analysisUtils task \texttt{au.estimateMRS} from the fifth percentile shortest baseline.}
\tablenotetext{d}{Data previously published in: I18: \citet{ilee18}, C17: \citet{cyganowski17}, I16: \citet{ilee16}}
\tablenotetext{e}{For the continuum image, except for ALMA 1.05 and 0.82\,mm, where it is for the line shown in Figure~\ref{fig:channelmaps}.  \emph{u,v}-ranges were used for the 1.3\,mm ($>$25\,k$\lambda$) and 3\,cm  ($>$1300\,m$\approx$43\,k$\lambda$) continuum images due to sparse sampling of shorter spacings and to minimize artifacts from the G11.94$-$0.62 HII region, respectively.}
\tablenotetext{f}{Measured near MM2.  Median values are quoted for line data ($\Delta$v is the channel width); the rms varies channel-to-channel due to variations in atmospheric opacity and bright and/or poorly imaged extended structures within the field of view \citep[see also][]{cyganowski17}.  }
\end{deluxetable*}

\begin{figure*}
    \centering
    \includegraphics[width=1.0\linewidth]{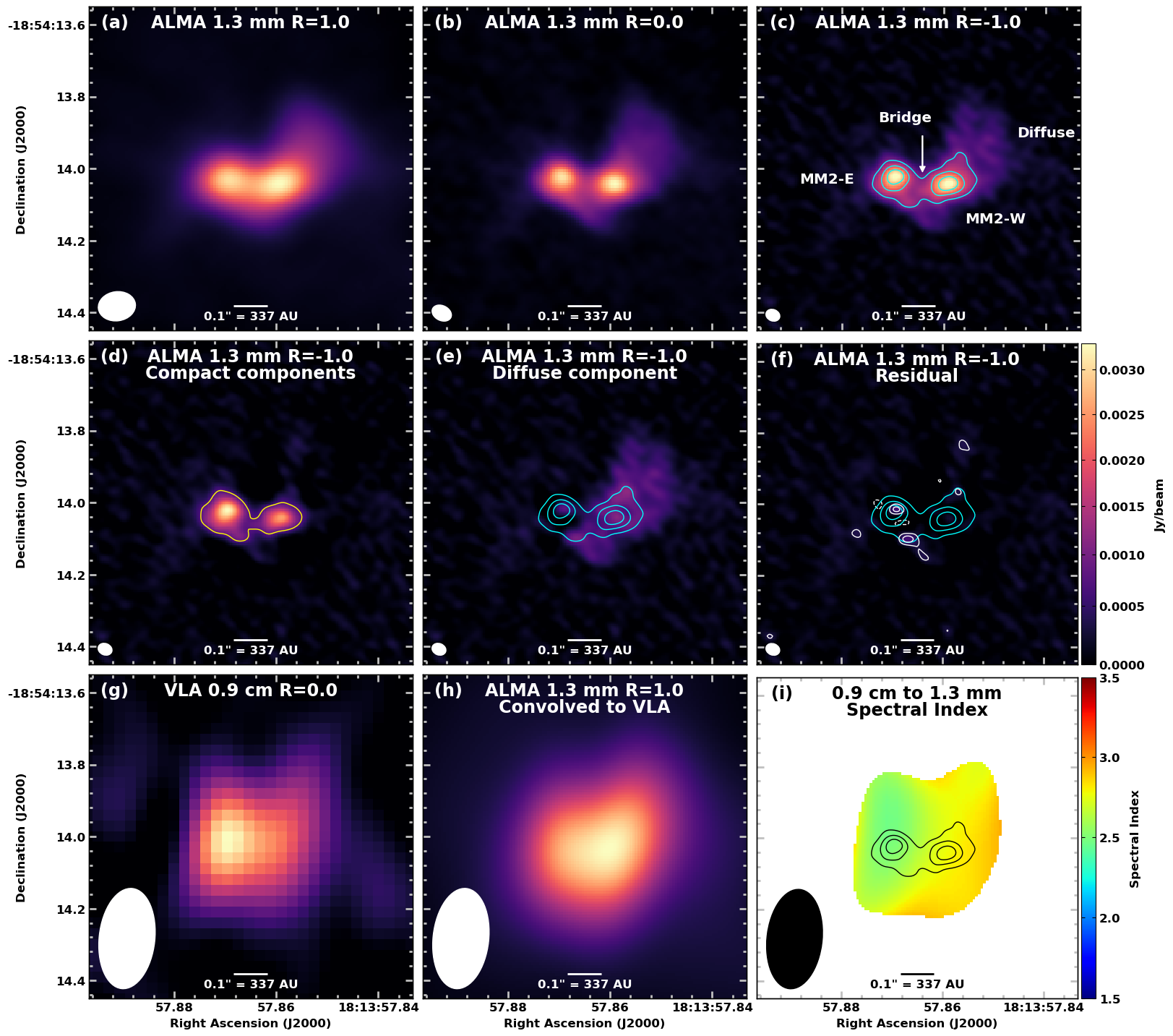}
    \caption{(a-c): ALMA 1.3~mm continuum images made with robust parameters of (a) R=1.0, (b) R=0.0, and (c) R=$-$1.0, with millimeter sources labelled in (c).  (d-f): R=$-$1.0 images of (d) the two compact components only, (e) the diffuse component only and (f) the residual of the simultaneous three-component fit (\S\ref{sec:cont_results}).  (d)\&(f) also show contours of the image shown in colourscale, levels: (d) [10]$\times\sigma$ (yellow), (f) [4,7]$\times\sigma$ (solid white), [$-$4]$\times\sigma$ (dashed white), $\sigma=$7.4e$-$5 \jyb.  (g-h): ALMA 1.3~mm (g) and VLA 0.9~cm (h) images convolved to a common beam (0\farcs280$\times$0\farcs155, PA=$-$6.2$^{\circ}$).  (i): spectral index image created from (g) and (h), masked at the 5$\sigma$ level.  Panels (d-f) are displayed using a common colorscale, shown to the right of (f); panels (a-c) and (g-h) similarly use a power-law stretch with exponent=0.9 and a minimum of 0.0, but with the maximum set to the peak value for MM2 of each individual image.
    Contours of the R=$-$1.0 Planck $T_b$ image are overplotted in cyan in (c) and (e-f) and in black in (i), levels: [25,40,55~K].  The synthesized beam is shown at lower left in each panel.  }
    \label{fig:contfig}
\end{figure*}

\section{Observations}\label{sec:obs}

Here we describe the new ALMA data presented in this Letter; for completeness, Table~\ref{tab:obstable} summarizes observational parameters for all datasets used in our analysis. Estimated absolute flux calibration uncertainties are 5\% for the ALMA and VLA 3\,cm data and 10\% for the VLA 0.9\,cm data.  All measurements were made from images corrected for the primary beam response. 

Our Cycle 6 1.3~mm ALMA observations (PI: Ilee) were calibrated using the ALMA science pipeline (CASA 5.6.1-6).  The approach described in \citet{brogan16,cyganowski17} was used to identify line-free channels and construct a pseudo-continuum dataset; the resulting aggregate continuum bandwidth is $\sim$0.72~GHz.  The continuum data were iteratively self-calibrated and the solutions applied to the line data.  
We combined these new C43-8 data with the C40-7 data from \citet{ilee18} taken with a nearly identical tuning.  Combined continuum images were made using multi-frequency synthesis, two Taylor terms (to account for the spectral index of the emission across the observed bandwidth), multi-scale clean and Briggs weighting with a range of values of the robust (R) parameter (see Table~\ref{tab:obstable}).  
The combined line data were imaged with R=0.5 and a common velocity resolution of 0.7\,\kms.
We estimate the absolute positional uncertainty of the combined images as 7.4\,mas.
As the 1.3\,mm ALMA pointings were centered on MM1, MM2 lies at the $\sim$83\% level of the primary beam  in these data.  

Our 0.82\,mm ALMA data (PI: Cyganowski) were calibrated using the ALMA science pipeline (CASA 5.4.0).
We similarly constructed a pseudo-continuum dataset (aggregate continuum bandwidth $\sim$0.44\,GHz), iteratively self-calibrated the continuum, and applied the solutions to the line data.
Here we consider only the CH$_3$OH 4$_{\rm -1,3}$--3$_{\rm 0,3}$ line
within the wide spectral window (spw) included in the tuning to provide continuum sensitivity; results for the targeted N$_2$H$^{+}$(4-3) line will be presented in a forthcoming publication (Zhang et al.\ in prep.).  The CH$_3$OH 4$_{\rm -1,3}$--3$_{\rm 0,3}$ line was imaged with a velocity resolution of 1.0\,\kms.

\begin{deluxetable*}{lccccccc}
\tablewidth{0pc}
\setlength{\tabcolsep}{0.5mm}
\tablecaption{Fitted Source Properties \label{tab:fittable}}  
\tablehead{ Source & \multicolumn{2}{c}{Position (J2000)\tablenotemark{a}} & Peak Intensity\tablenotemark{a} & Integ. flux\tablenotemark{a} & T$_b$\tablenotemark{b} & Size\tablenotemark{a} & Size\tablenotemark{a}  \\
\colhead{} & \colhead{$\alpha$ ($^{\rm h}$ $^{\rm m}$ $^{\rm s}$)} & \colhead{$\delta$ ($^{\circ}$ \arcmin~\arcsec)} & \colhead{(\mjb)} & \colhead{density (mJy)}  & {(K)} & {(\arcsec$\times$\arcsec [P.A.($^{\circ}$)])} & (au$\times$au)
}
\startdata
\multicolumn{2}{l}{\textbf{ALMA 1.3~mm R=-1.0}}& & & & & & \\
\MME &  18:13:57.86993 & $-$18:54:14.0305 & 2.91 (0.07) &  17.9 (0.5) & 68.4 & 0.085$\times$0.078 (0.003) [$+$55 (21)] & 286$\times$262 (9) \\
\MMW &  18:13:57.85941 & $-$18:54:14.0445 &  2.42 (0.07) &  10.8 (0.4) & 64.6 &  0.088$\times$0.048 (0.004) [$+$111 (3)] & 295$\times$163 (12)\\
Diffuse &  18:13:57.8570 &  $-$18:54:14.006 & 0.94 (0.03) & 39 (1) & 22.7 &   0.305$\times$0.169 (0.01) [$+$135 (2)] & 1029$\times$570 (35) \\
\multicolumn{2}{l}{\textbf{VLA 0.9~cm R=0.0}\tablenotemark{c}}& & & & & & \\
\MME & fixed & fixed & 0.139 (0.008) & 0.14 (0.02) & 4.9 & $<$0.270$\times<0.144$  & $<$910$\times<$485\\
\MMW$+$diffuse & fixed & fixed & 0.094 (0.009) & 0.21 (0.03) & 6.3 & 0.213$\times$0.197 (0.09) [$+$83 (61)] & 718$\times$664 (300) \\
\enddata
\tablenotetext{a}{Properties from 2D Gaussian fitting (\S\ref{sec:cont_results}): ``size'' is the FWHM deconvolved source size, statistical uncertainties are given in parentheses or indicated by the number of significant figures.      }
\tablenotetext{b}{Planck $T_b$ calculated from $S_\nu$ and FWHM fitted size.}
\tablenotetext{c}{Positions fixed to those of \MME/\MMW\/ from the 1.3\,mm fit.  For \MME, the beamsize is used in calculating T$_b$ and reported as an upper limit for the size, as the source could not be deconvolved from the beam.}
\end{deluxetable*}

\section{Results}

\subsection{ALMA 1.3 mm Continuum Emission}
\label{sec:cont_results}

Figure~\ref{fig:contfig} shows our ALMA 1.3~mm continuum images of G11.92$-$0.61 MM2.  The most striking feature of these high resolution images (beam $\lesssim$160\,au; Figure~\ref{fig:contfig}b,c) is that the 1.3~mm continuum is clearly resolved into two compact sources, which we designate \MME\/ and \MMW.  These two compact sources are connected by a ``bridge'' of lower surface brightness emission; diffuse, low-surface-brightness emission also extends N/NW of \MMW\/ (labelled ``Diffuse'' in Figure~\ref{fig:contfig}c) and to the south of the connecting bridge.

To characterize the properties of the compact sources, we fit the  R=$-$1.0 1.3~mm continuum image with two-dimensional Gaussians.  Three components are required to
represent the emission: one each for \MME\/ and \MMW\/ and a third, more extended component for the diffuse emission.  
The fitted properties of these components are given in Table~\ref{tab:fittable} and the fitting results are illustrated by Figure~\ref{fig:contfig}d-f.  
Notably, the residual image contains an $\sim$8.1$\sigma$ peak coincident with \MME\/ (0\farcs017$\sim$57\,au N/NW of its fitted position), indicating that this source is not entirely Gaussian.  There is also an $\sim$8.5$\sigma$ peak 0\farcs083 ($\sim$280\,au) S/SW of \MME.  Both residuals suggest the existence of further substructure, including possible further multiplicity unresolved by our observations.

The projected separation between \MME\/ and \MMW\/ is 0\farcs1499$\sim$505\,au.
Their connecting ``bridge'', detected with 10$\sigma<\frac{S}{N}<$11$\sigma$ in  the compact-component-only image (Figure~\ref{fig:contfig}d), has a width of $\sim$0\farcs03$\sim$100\,au, estimated from the 10$\sigma$ contour.
The compact sources 
lie within larger structure(s), as shown by the differences in the R=1,0,$-$1 images (Figure~\ref{fig:contfig}a-c) and the need for a diffuse component in fitting the R=$-$1 image.  
Using CASA's \textsc{imstat} task, we estimate the integrated flux density ($S_\nu$) of $>$4$\sigma$ emission as 
$\approx$107($\pm6$), 77($\pm$3), and 58($\pm$5)\,mJy for the R=1,0,$-$1 images, respectively \citep[uncertainties estimated following][]{cyganowski12}.
As expected, more extended emission is also detected in the lower-resolution images: the R=1 image recovers filamentary emission extending $\sim$1\farcs5 (5000\,au) N/NW of \MMW\/ (beyond the field of Figure~\ref{fig:contfig}, see also \S\ref{sec:lines_outflow}) 
and
the E-W extent of $>$4$\sigma$ emission around
\MME/\MMW\/ is $\sim$1\farcs2, 0\farcs8, and 0\farcs4 in the R=1,0,$-$1 images. 
Even in the R=$-$1 image,
the diffuse component accounts for 58$\pm$4\% of the fitted integrated fitted flux density (Table~\ref{tab:fittable}).

\subsection{Spectral Index and VLA cm Continuum Emission}
\label{sec:alpha}

To constrain the spectral indices ($\alpha$) of \MME\/ and \MMW, we combine our new ALMA 1.3\,mm images with previously published 0.9\,cm and 3\,cm VLA data (Table~\ref{tab:obstable}).  To achieve the best compromise between angular resolution and sensitivity, we reimaged the 0.9\,cm data with R=0 \citep[using two Taylor terms and multifrequency synthesis, as described in][]{ilee16}.  The emission is elongated E-W \citep[Figure~\ref{fig:contfig}g, see also][]{hunter15} with a morphology consistent with two sources only marginally resolved.
Notably, at 0.9\,cm the eastern source is brighter, while at 1.3\,mm the western source is brighter (Figure~\ref{fig:contfig}a-c,h), although the latter includes contributions from \MMW\/ and the diffuse component discussed in \S\ref{sec:cont_results}.  To visualize the variation in spectral index across MM2, Figure~\ref{fig:contfig}i shows the \myalpha\/ image calculated from the images in Figure~\ref{fig:contfig}g,h: \myalpha\/ ranges from 2.47--2.94, being lower to the east.
MM2 is undetected ($<$4$\sigma$) in the 3\,cm VLA image.
To estimate 3\,cm upper limits for \MME\/ and \MMW, we measure the peak intensity of the 3\,cm emission within the 10\% contour of the 0.9\,cm emission, yielding
$<$19.1\,\ujb\/ ($\sim$3.4$\sigma$). 

The \myalpha\/ values and 3\,cm nondetections of \MME\/ and \MMW\/ indicate
partially optically thick thermal dust emission.
Extrapolating the $S_\nu$ of \MME\/ and \MMW\/ from the 1.3\,mm R=$-$1.0 image
(Table~\ref{tab:fittable}) to 3\,cm 
using the shallowest observed \myalpha=2.47 predicts 
S$_{\rm 3\,cm}$=8$\pm$1 and 4.7$\pm$0.9\,$\mu$Jy
respectively, consistent with our 3\,cm nondetections.
As an additional check, we fit the 0.9\,cm image with two 2D Gaussian components, fixing their positions to those of \MME\/ and \MMW\/ from \S\ref{sec:cont_results} and noting that in the lower-resolution 0.9\,cm image, the western component represents a combination of the compact source \MMW\/ and diffuse emission.  
Extrapolating these fitted 0.9\,cm flux densities (Table~\ref{tab:fittable}) predicts $S_{\rm 3\,cm}$=8$\pm$4 and 11$\pm$8\,$\mu$Jy, again consistent with our 3\,cm nondetections.

\begin{figure*}
    \centering
    \includegraphics[width=0.85\linewidth]{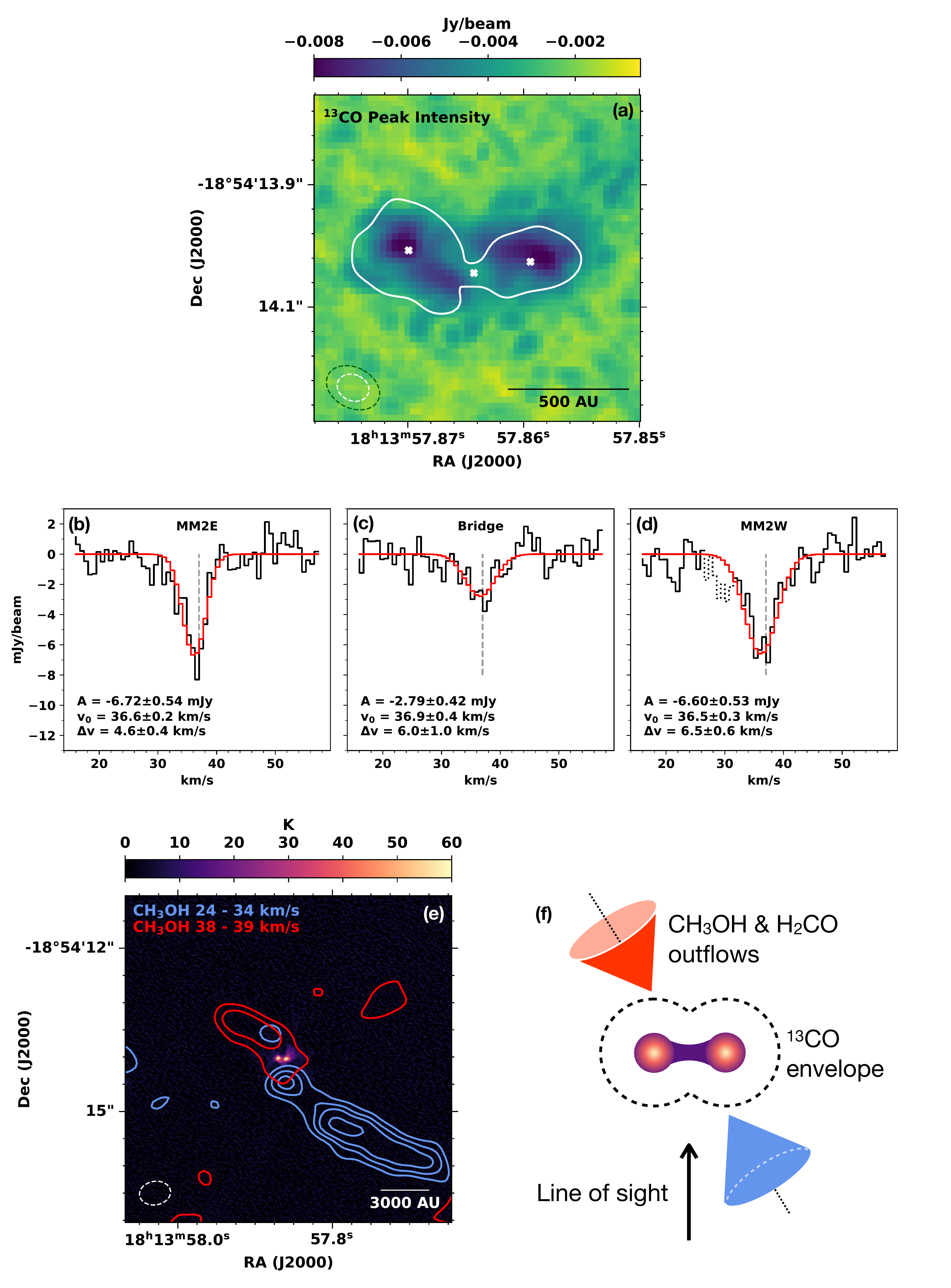}
    \caption{(a): Zoom of MM2 showing minimum map of $^{13}$CO J=2-1 absorption.  The white 10$\sigma$ continuum contour from Figure~\ref{fig:contfig}d shows the ``bridge'' (\S\ref{sec:cont_results}).  Crosses mark
    pixels for which spectra are shown. 
    (b-d): $^{13}$CO J=2-1 spectra (black) at 
    the fitted continuum positions of \MME\/ and \MMW\/ and a 
    ``bridge'' pixel, overplotted with Gaussian fits 
    to the line core 
    (red; the \MMW\/ fit excludes the channels shown with a dotted line).  The estimated systemic velocity (37 \kms, \S\ref{sec:lines_outflow}) is shown as a dashed gray line; labels give best-fit parameters.
    (e): 1.3\,mm Planck $T_b$ (R=$-1$) image overlaid with contours of integrated red/blueshifted CH$_3$OH 4$_{\rm -1,3}$--3$_{\rm 0,3}$ emission.  Levels: [4,7]$\times\sigma=$3.3\,\mjb\,\kms\/ (red), [4,7,10,15]$\times\sigma=$5.5\,\mjb\,\kms\/ (blue).
    (f): Proposed morphology of MM2, viewed perpendicular to the line of sight.  (a)\&(e): ALMA synthesized beams are shown at bottom left.}
    \label{fig:13co}
\end{figure*}

\subsection{Line Absorption from the Compact Core}
\label{sec:lines_core}

To identify molecular gas potentially associated with the compact millimeter continuum sources, we searched for $\ge$4$\sigma$ emission or absorption that spanned $\ge$2 adjacent channels in the combined 1.3\,mm line cubes at the positions of \MME\/ and \MMW\/ (Table~\ref{tab:fittable}). 
The only line detected in these high-resolution data (beam$\lesssim$0\farcs08$\approx$270 au) is $^{13}$CO J=2-1 ($\nu_{\rm rest}=$220.398684 GHz, E$_{\rm upper}$=16\,K), in absorption against the 
1.3\,mm continuum.

To study the $^{13}$CO absorption, we imaged this line with R=0, a compromise between spatial resolution and signal-to-noise ratio (SNR) that yields a synthesized beamsize of 0\farcs069$\times$0\farcs050 ($\approx$230$\times$170 au).
Figure~\ref{fig:13co}a, a map of the minimum value of the $^{13}$CO spectrum at each pixel, illustrates the spatial morphology of the 
absorption.  Notably, absorption extends across \MME, \MMW\/ and 
the continuum ``bridge'' 
but the depth of the absorption does not exactly follow the strength of the 1.3\,mm continuum.
The deepest absorption towards \MME\/ is 0\farcs006 (20\,AU) east of the strongest continuum emission, while towards \MMW\/ the deepest absorption is 0\farcs008 (27\,AU) northwest of the continuum peak.
Gaussian fitting of the absorption profiles (Figure~\ref{fig:13co}b-d) yields similar centroid velocities for \MME\/ and \MMW, while \MMW\/ has a broader linewidth ($\Delta$v$_{\rm FWHM}=4.6\pm0.4$ and $6.5\pm0.6$\,km\,s$^{-1}$, respectively). 
Taken together, these
results 
suggest the $^{13}$CO absorption traces gas 
physically associated with \MME\/ and \MMW, rather than e.g.\ a 
line-of-sight foreground cloud. 
This is reminiscent of the scenario outlined in \citet{sahu19} where an absorbing layer of cooler molecular material surrounds the optically thick protostar and binary member NGC1333-IRAS4A1, which exhibits a millimeter dust $T_b$ ($\sim$60\,K) similar to \MME\/ and \MMW.

\subsection{Outflow Line Emission}
\label{sec:lines_outflow}

\begin{figure*}
    \centering
    \includegraphics[width=1.0\linewidth]{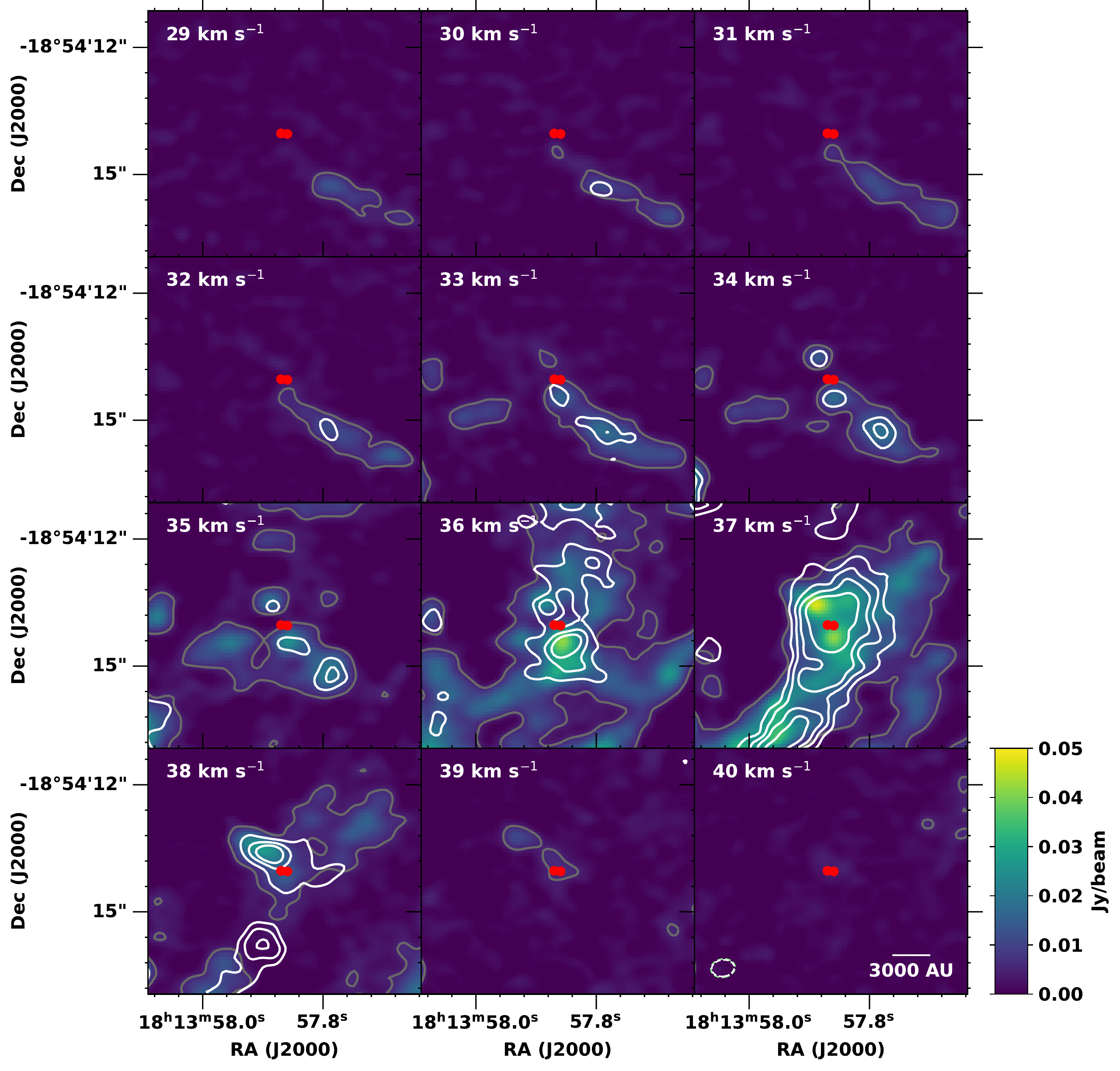}
    \caption{Channel maps showing CH$_3$OH 4$_{\rm -1,3}$--3$_{\rm 0,3}$ (colorscale and grey contour, 4$\times\sigma=$1.2 \mjb) and H$_2$CO 4$_{\rm 0,4}$--3$_{\rm 0,3}$ (white contours, [4,7,10,15]$\times\sigma=$4.9 \mjb) emission.  Red circles mark the positions of \MME\/ and \MMW\/ (Table~\ref{tab:fittable}).   }
    \label{fig:channelmaps}
\end{figure*}

Molecular outflows are clear signposts of protostars that have revealed star formation activity in other candidate high-mass starless cores \citep[e.g.][]{duartecabral13, tan16, Pillai19}.
While $^{13}$CO is not detected in emission near MM2 in our high-resolution data, low-excitation lines of CH$_3$OH and H$_2$CO provide alternative tracers of outflows from low- and high-mass protostars \citep{Brogan09,Tychoniec21,Morii21}. Fortuitously, the tuning and larger LAS of our deep 0.82\,mm ALMA observations (Table~\ref{tab:obstable}) provide an opportunity to search for outflow activity from \MME/\MMW\/ using CH$_3$OH 4$_{\rm -1,3}$--3$_{\rm 0,3}$ 
($\nu_{\rm rest}$=358.605799 GHz, E$_{\rm upper}$=44\,K).
Figure~\ref{fig:channelmaps} shows channel maps of this CH$_3$OH line, illustrating that on the larger scales probed by these data (beam 0\farcs50$\approx$1685 au) MM2 lies on a filament aligned roughly N-S \citep[see also][]{cyganowski17}.  At the positions of \MME\/ and \MMW, the CH$_3$OH 
emission from the filament peaks at $\sim$37 \kms.   
Taking this estimate of MM2's systemic velocity,  blueshifted CH$_3$OH emission extends southwest of \MME/\MMW, while redshifted CH$_3$OH emission lies to the northeast  (Figure~\ref{fig:13co}e,\ref{fig:channelmaps}).
This kinematic morphology 
suggests an asymmetric bipolar molecular outflow driven by \MME\/ and/or \MMW.
The projected length and velocity extent are $\sim$12,600\,au and 13\,\kms\/ for the blueshifted
lobe and $\sim$5,100\,au and 2\,\kms\/ for the redshifted lobe (lengths are the average of estimates assuming the driving source is \MME/\MMW).
These values imply dynamical timescales of t$_{\rm dyn}$$\sim$4,600 years and $\sim$12,100 years for the blue and red lobes.  
We emphasize, however, that 
the CH$_3$OH emission is unlikely to trace the highest-velocity gas \citep[see $^{12}$CO/H$_2$CO comparison for a low-mass outflow in][]{cyganowski17} so
these t$_{\rm dyn}$ estimates should be interpreted with caution.

To check for evidence of this outflow in other lines, we reimaged the five H$_2$CO and CH$_3$OH transitions with E$_{\rm upper}$$<$100\,K in the 1.05\,mm tuning of \citet{cyganowski17}.  Figure~\ref{fig:channelmaps} shows H$_2$CO 4$_{0,4}-$3$_{0,3}$ ($\nu_{\rm rest}$=290.62341\,GHz, E$_{\rm upper}$=35\,K), the closest to CH$_3$OH 4$_{\rm -1,3}$--3$_{\rm 0,3}$ in  E$_{\rm upper}$ and line strength and the only 1.05\,mm line observed with sufficient spectral resolution to image with $\Delta$v=1\,\kms\/ (Table~\ref{tab:obstable}; the others were imaged with $\Delta$v=1.2\,\kms).  
The behavior of this H$_2$CO line is representative of the 1.05\,mm H$_2$CO and CH$_3$OH transitions, with similar morphology to CH$_3$OH 4$_{\rm -1,3}$--3$_{\rm 0,3}$ in channels near the systemic velocity (Figure~\ref{fig:channelmaps}), but outflow emission detected over a narrower velocity range and at lower SNR due to the lower sensitivity of the data (Table~\ref{tab:obstable}).  Figure~\ref{fig:13co}f summarizes the proposed morphology of the 
core/outflow 
system.

\section{Discussion}

To explore the nature of \MME\/ and \MMW,
we first consider their observed 1.3~mm continuum 
brightness temperatures ($T_b$), which  provide strict lower limits for their physical temperatures of 68.4 K and 64.6 K respectively (Table~\ref{tab:fittable}).  These high temperatures signify internal heating, as external heating 
\citep[by MM1 and the intermediate or high-mass protostar MM3;]
[]{cyganowski09,cyganowski11a,cyganowski17} could account for dust temperatures of at most $\sim$23 K, based on simple estimates \citep[see also][]{cyganowski14}.
With evidence for both internal heating and a bipolar outflow (\S\ref{sec:lines_outflow}), we interpret \MME\/ and \MMW\/ as deeply embedded protostars, which leads to the conclusion that MM2 is not starless and emphasizes the importance of high-resolution (sub)millimeter observations for detecting protostars and their outflows in candidate high-mass starless clumps and cores \citep[see also e.g.][]{duartecabral13,tan16,Pillai19,svoboda19}.

Observed (sub)millimeter $T_b$ 
can be used to estimate the total luminosities ($L_*$) of deeply embedded protostars in the context of a simple model of blackbody emission from an optically thick dust shell surrounding them 
\citep[e.g.][]{brogan16,hunter17,ginsberg17}, 
via
\begin{math}
    L_* = 4 \pi r^2 \sigma T_b^4,
\end{math}
where r is the radius of the $\tau\approx1$ sphere and $\sigma$ is the Stefan-Boltzmann constant.
Because the observed dust emission is not entirely optically thick 
toward \MME\/ and \MMW\/ (\S\ref{sec:alpha}, Figure~\ref{fig:contfig}i), the observed $T_b$ will underestimate the dust
temperature and $L_*$ will be a lower limit.  Using their fitted sizes and Planck $T_b$  
(calculated from the integrated flux densities and fitted sizes; Table~\ref{tab:fittable}), we estimate L$_*>$24.7 \lsun\/ for \MME\/ and L$_*>$12.6 \lsun\/ for \MMW.
Notably, these limiting values are 1-4 orders of magnitude higher than those estimated 
for the low-mass members of the NGC6334I protocluster using the same approach \citep[MM5-9, Table 5 of][]{brogan16}.
Although our limiting luminosities for \MME/\MMW\/ are $\sim$3 orders of magnitude lower than those estimated with this method
for W51e2e and for NGC6334I-MM1 in outburst \citep[2.3$\times$10$^4$ \lsun\/ and 4.2$\times$10$^4$ \lsun, respectively;][]{ginsberg17,hunter17}, massive protostars are expected to pass through a low-luminosity stage early in their evolution \citep[e.g.][]{kuiperyorke13}.

The closest analogue to \MME\/ and \MMW\/ in the literature is
NGC6334I-MM4A, 
an optically thick dust source, in a massive protocluster, that lacks compact thermal molecular line emission in ALMA observations despite a high dust $T_b$ \citep[97$\pm$5 K;][]{brogan16}.  NGC6334I-MM4A drives a collimated bipolar outflow detected in dense gas tracers and exhibits faint, variable water maser emission \citep{Brogan18}.
Although 
previous surveys found no water masers toward MM2 
\citep{HC96,Breen11}, masers with luminosity similar to those in NGC6334I-MM4A would be only $\sim$0.05 Jy at G11.92$-$0.61's distance, and would have been undetected by these surveys, particularly at velocities where the bright MM1 maser limits the image dynamic range. 
An analogue in a low-mass multiple system is component B of IRAS\,16293-2422, a partially optically thick dust source with T$_{b,\,mm}\sim$180~K, interpreted as a very young protostar \citep{chandler05,hernandez19}.
To our knowledge, \MME/\MMW\/ is the first example 
of a system of two nearly optically thick millimeter dust sources.

Comparing our results with model predictions 
(\S\ref{sec:intro}), 
the observed properties of \MME\/ and \MMW\/ in many respects match the super-Alfv\'{e}nic cases of \citet{mignon-risse21} remarkably well, suggesting that this protobinary may be forming in an environment with a weak magnetic field.  
\MME\/ and \MMW\/
have similar 1.3\,mm $T_b$ and $S_\nu$ (ratio E:W=1.06 and 1.66, respectively),
suggesting that the mass ratio of the two protostars is likely comparable to the $\approx$1.1-1.6 range 
of the \citet{mignon-risse21} simulations.
The observed separation of \MME\/ and \MMW\/ ($\sim$505\,au)
is similarly consistent with the \citet{mignon-risse21} results (binary separations 350-700\,au) 
and the ``bridge'' we observe 
is qualitatively similar to 
linking structures visible in 
the simulated 
column density 
maps 
in their Figure 8. 
Interestingly, linking ``bridges'' form in simulations of binary formation via both core \citep[e.g.][for $\sim$equal-mass low-mass binaries]{riaz14} and disk fragmentation \citep[e.g.][in which disk fragmentation is precipitated by the collision of extended spiral arms]{mignon-risse21}.

In the \citet{mignon-risse21} simulations, the individual protostars have Keplerian disks with diameters of $\sim$200-400\,au, which are embedded within a transient disk-like 
circumbinary structure.
With no detected line emission in our high-resolution observations, 
it is unclear
whether \MME\/ and \MMW\/ exhibit Keplerian rotation. 
Their fitted sizes (Table~\ref{tab:fittable}) are, however, comparable to the simulation's individual disk diameters, with the more extended millimeter emission 
potentially tracing a circumbinary 
disk.
The total circumprotostellar gas mass from the sum of the fitted VLA 0.9\,cm flux densities is 6.8$\pm$0.9 \msun\/ (considering the fitting uncertainties from Table~\ref{tab:fittable}, added in quadrature, and 10\% calibration uncertainty)
using
T$_{dust}=$66.5~K (the average 1.3\,mm T$_b$ for \MME/\MMW) 
and, 
following the approach of \citet{Karnath20}, assuming
the 0.9\,cm emission is optically thin, a gas:dust mass ratio of 100:1, and $\kappa_{\rm 0.9\,cm}=$0.128\,cm$^2$\,g$^{-1}$.

Notably, this estimate is comparable to the sum of the virial masses calculated from the $^{13}$CO linewidths (\S\ref{sec:lines_core}): assuming spherical clouds with 1/r density profiles \citep{Carpenter90}, angular diameters equal to the geometric means of the 1.3\,mm fitted sizes (Table~\ref{tab:fittable}), and correcting for a mean inclination of the rotation axis to the line of sight (30$^{\circ}$) yields 3.5$\pm$0.8\,M$_{\odot}$ for \MME\/ and 5.7$\pm$1.0\,M$_{\odot}$ for \MMW\/ for a total of 9.2$\pm$1.2\,M$_{\odot}$.
To test the dependence 
on the assumed angular diameter, we used a similar fitting procedure to obtain source sizes from the R=0 1.3\,mm continuum image, which yields a combined virial mass of 9.0$\pm$1.2\,M$_{\odot}$. 
The combined virial mass of $\approx$9$\pm$1\,M$_{\odot}$ minus the gas mass estimate allows for central protostars of current mass $\sim$1\,M$_{\odot}$.  Depending on their evolutionary track, the $L_*$ (including accretion) of such protostars can reach values of $\sim$25\,L$_{\odot}$ 
\citep{Young05} to $>$10$^{3}$\,L$_{\odot}$ \citep{kuiperyorke13},  consistent with the $L_*$ lower limits that we derive from the dust $T_b$ of \MME\/ and \MMW\/.
Considering the luminosity limits, protostellar mass estimates and dust properties derived above together with theoretical expectations, we interpret \MME\/ and \MMW\/ as a young proto-high-mass-binary system.

The outflow from \MME/\MMW\/ (\S\ref{sec:lines_outflow}, Figure~\ref{fig:13co}e\&\ref{fig:channelmaps}) provides evidence for ongoing accretion onto the growing protobinary system.  With ample fuel available within the gas-rich protocluster environment, the protostellar masses (and luminosities) are expected to increase with time.  Accretion will also affect the binary separation, which can increase or decrease 
depending on turbulence, magnetic field strength, and the presence of outflows, with magnetic fields 
promoting 
the formation of close high-mass binary systems \citep[e.g.][]{lund18,harada21,Ramirez21}.  
Future high-resolution observations of \MME/\MMW\/ -- at shorter wavelengths to better measure the protostellar luminosities, at longer wavelengths to search for line emission in a regime where the dust is optically thin, and in full polarization to measure the magnetic field --
will provide a powerful test case for models of high-mass binary formation.

\acknowledgments
The National Radio Astronomy Observatory is a facility of the National Science Foundation operated under cooperative agreement by Associated Universities, Inc. This paper makes use of the following ALMA data: ADS/JAO.ALMA\#2013.1.00812.S,
ADS/JAO.ALMA\#2015.1.00827.S,\\ 
ADS/JAO.ALMA\#2016.1.01147.S,\\
ADS/JAO.ALMA\#2017.1.01373.S, and\\ ADS/JAO.ALMA\#2018.1.01010.S. ALMA is a partnership of ESO (representing its member states), NSF (USA) and NINS (Japan), together with NRC (Canada), MOST and ASIAA (Taiwan), and KASI (Republic of Korea), in cooperation with the Republic of Chile. The Joint ALMA Observatory is operated by ESO, AUI/NRAO and NAOJ.  C.J.C. acknowledges support from the University of St Andrews Restarting Research Funding Scheme (SARRF), which is funded through the SFC grant reference SFC/AN/08/020.  J.D.I. acknowledges support from the UK's STFC under ST/T000287/1. S.Z. is funded by the China Scholarship Council--University of St Andrews Scholarship (PhD programmes, No. 201806190010).  T.J.H. is funded by a Royal Society Dorothy Hodgkin Fellowship. This research made use of NASA's Astrophysics Data System Bibliographic Services and APLpy, an open-source plotting package for Python \citep{aplpy2012}.

\bibliography{bibliography}


\end{CJK*}
\end{document}